\documentclass[prl,twocolumn,superscriptaddress,floatfix,nopacs,longbibliography]{revtex4-1}
\usepackage[utf8]{inputenc}

\usepackage{amsmath}
\usepackage{amssymb}
\usepackage{times}
\usepackage{tikz}
\usepackage{booktabs}
\usepackage{mathtools}
\usepackage{multirow}
\usepackage{natbib}
\usepackage{siunitx}
\usepackage[normalem]{ulem}

\usepackage{standalone}
\usepackage{braket}

\begin{document}

\title{Finite temperature tensor network algorithm for frustrated two-dimensional quantum materials}
\author{Philipp Schmoll}
\affiliation{Dahlem Center for Complex Quantum Systems and Institut f\"{u}r Theoretische Physik, Freie Universität Berlin, Arnimallee 14, 14195 Berlin, Germany}
\author{Christian Balz}
\affiliation{ISIS Neutron and Muon Source, Rutherford Appleton Laboratory, Didcot OX11 0QX, UK}
\author{Bella Lake}
\affiliation{Helmholtz-Zentrum Berlin f\"{u}r Materialien und Energie, Hahn-Meitner-Platz 1, 14109 Berlin, Germany}
\affiliation{\mbox{Institut f\"{u}r Festk\"{o}rperphysik, Technische Universit\"{a}t Berlin, Hardenbergstra{\ss}e 36, D-10623 Berlin, Germany}}
\author{Jens Eisert}
\affiliation{Dahlem Center for Complex Quantum Systems and Institut f\"{u}r Theoretische Physik, Freie Universität Berlin, Arnimallee 14, 14195 Berlin, Germany}
\affiliation{Helmholtz-Zentrum Berlin f\"{u}r Materialien und Energie, Hahn-Meitner-Platz 1, 14109 Berlin, Germany}
\author{Augustine Kshetrimayum}
\affiliation{Dahlem Center for Complex Quantum Systems and Institut f\"{u}r Theoretische Physik, Freie Universität Berlin, Arnimallee 14, 14195 Berlin, Germany}
\affiliation{Helmholtz-Zentrum Berlin f\"{u}r Materialien und Energie, Hahn-Meitner-Platz 1, 14109 Berlin, Germany}
\affiliation{Theory Division, Saha Institute of Nuclear Physics, 1/AF Bidhannagar, Kolkata 700 064, India}

\begin{abstract}

Aimed at a more realistic classical description of natural quantum systems, we present a two-dimensional tensor network algorithm to study finite temperature properties of frustrated model quantum systems and real quantum materials. For this purpose, we introduce the infinite projected entangled simplex operator ansatz to study thermodynamic properties. To obtain state-of-the-art benchmarking results, we explore the highly challenging spin-1/2 Heisenberg anti-ferromagnet on the Kagome lattice, a system for which we investigate the melting of the magnetization plateaus at finite magnetic field and temperature. Making close connection to actual experimental data of real quantum materials, we go on to studying the finite temperature properties of Ca$_{10}$Cr$_7$O$_{28}$. We compare the magnetization curve of this material in the presence of an external magnetic field at finite temperature with classically simulated data. As a first theoretical tool that incorporates both thermal fluctuations as well as quantum correlations in the study of this material, our work contributes to settling the existing controversy between the experimental data and previous theoretical works on the magnetization process.
\end{abstract}

\maketitle

\section{Introduction}

Simulating complex quantum materials is considered to be one of the hardest problems in modern physics. Density functional theory is arguably the most popular approach to date for calculating the electronic structure of molecules and extended materials~\cite{PhysRev.140.A1133,PhysRevLett.102.073005}. In situations in which strong correlations are expected to be dominant, however, its applicability can be limited. Ultimately, the core computational challenge in the numerical simulation of strongly correlated quantum materials arises from the exponential scaling of the size of the Hilbert space with the system size. Thus, it comes as no surprise that the \emph{exact diagonalization} (ED) technique can only study small sizes and therefore may fail to capture the important physics of emergent many-body phenomena. Mean-field techniques are also unsuitable in the study of quantum materials as they neglect the most crucial ingredient in describing these systems: quantum entanglement. While \emph{quantum Monte Carlo} constitutes a versatile tool for simulating unfrustrated strongly correlated systems~\cite{QMC}, they suffer from severe limitations for frustrated quantum systems due to the sign problem. In this respect, \emph{tensor network} techniques have emerged as a powerful alternative for studying challenging many-body problems which does not suffer from any of those limitations~\cite{Orus-AnnPhys-2014,VerstraeteBig,Handwaving,AreaReview}.

The success of one-dimensional tensor networks, also known as \emph{matrix product states} (MPS)~\cite{dmrgwhite1992,dmrgRMP,Schollwoeck2011,MPSReps}, in describing one-dimensional phases of matter have provided much impetus to the development of two-dimensional tensor network algorithms. While the situation is much more intricate and challenging in two spatial dimensions, such tensor network algorithms, also known as \emph{projected entangled pair states} (PEPS) or
\emph{iPEPS}~\cite{Verstraete-arxiv-2004,iPEPSOld,VerstraeteBig} in its infinite instance tackling directly the thermodynamic limit, have recently matured and have been employed successfully to study various challenging problems in two dimensions. This includes finding ground states of frustrated systems and real quantum materials~\cite{Xiangkhaf,ThibautspinS,Thibautnematic,KshetrimayumkagoXXZ,Corbozmaterial,Kshetrimayummaterial,IqbalShuriken2021} and non-equilibrium systems~\cite{KshetrimayumNatcomm2017,Czarnikevolution2019,Hubig2019,Kshetrimayum2DMBL,Kshetrimayum2DTC,DziarmagaNhoodupdate2021,DziarmagaGradupdate2022PRB}. While most of the efforts has been dedicated towards identifying ground states of closed quantum systems, in order to accurately capture the physics of quantum materials in realistic conditions in the lab, one needs to include the effects of temperature. With this aim, there have been several recent works on two-dimensional \emph{finite temperature tensor network algorithms}~\cite{CzarnikfinT2012,CzarnikfinT2015,Kshetrimayum2019,CzarniKitaevfinT,CzarnikSSlandfinT,Mondal2020}. Most of these works have, however, focused on paradigmatic, theoretical models such as the Ising, Kitaev or Heisenberg models, and mostly models that are defined on the square lattice. 

In this work, we develop a two-dimensional tensor network algorithm for studying finite temperature properties of existing quantum materials, thus mimicking experimental studies as closely as possible. We start by describing our method and then present results on two important instances of strongly correlated systems: (i) the paradigmatic spin-$1/2$ Kagome Heisenberg anti-ferromagnet both in the absence and presence of an external magnetic field and (ii) the real quantum material Ca$_{10}$Cr$_7$O$_{28}$ that features a bilayer Kagome structure.

\section{Method}

Our method substantially advances the algorithm proposed in Ref.~\cite{Kshetrimayum2019} by extending it to the more challenging realm of frustrated quantum systems and real quantum materials. This step renders it possible to directly compare experimental data and theoretical tensor network simulations, as we do here. We will now review the underlying annealing algorithm and highlight the improvements. In order to simulate a quantum system at finite temperature $\beta \coloneqq 1 / T > 0$, we describe it by an (unnormalized) thermal quantum state
\begin{align}
    \rho(\beta) = \mathrm e^{-\beta H},
\end{align}
where $ H$ is the full local many-body Hamiltonian. To obtain such a Gibbs state, we start from an infinite temperature state, i.e., $\rho(\beta = 0)$ and cool down the system to the desired temperature $\beta^{-1}>0$. The initial state is simply a tensor product of identities, the (unnormalized) single-particle thermal state in the limit $T \rightarrow \infty$. The evolution to the desired temperature can be generated by suitably many small temperature steps $\delta\beta$, so that the full quantum state is obtained for $N\in \mathbb{N}$ by
\begin{align}
    \rho(\beta) = \rho(\delta\beta)^N = \left( \mathrm{e}^{-\delta\beta H} \right)^N
\end{align}
with $\delta\beta \coloneqq \beta / N$ and $\rho(\delta\beta)$ as what we call 
the \emph{infinitesimal thermal density matrix} (ITDM). This cooling is implemented by a simple update technique~\cite{simpleupdatejiang,Kshetrimayum2019}. The simple update is adopted here for its numerical stability and efficiency~\cite{Xiangkhaf,Thibautnematic,Kshetrimayum2019,GauthePRL2022}, particularly relevant while working on systems with large physical dimensions, which seems a necessity for the demanding task considered here (see the Appendix).

Instead of directly cooling down to $\beta>0$, it is advantageous to cool down to $\beta/2$ and evaluate the Gibbs state as
\begin{align}
    \rho(\beta) = \rho(\beta/2)^\dagger \rho(\beta/2)\ .
    \label{eq:doubeLayerThermalDensityMatrix}
\end{align}
This ensures that the resulting operator is \emph{positive semi-definite} and hence reflects a valid quantum state, which is otherwise not guaranteed in tensor network implementations due to truncation effects~\cite{Chen2018,PositiveMPO,Chen2018_2}. Eq.~\eqref{eq:doubeLayerThermalDensityMatrix} is \emph{the} main difference to the underlying algorithm presented in Ref.~\cite{Kshetrimayum2019} and is the crucial improvement which enables the simulation of frustrated systems (along with using the correct tensor network structure of the underlying lattice of the model as we discuss in the next paragraph). Thus, we have the freedom of evolving up to only $N/2$ steps thereby saving a factor of two in the number of annealing steps or evolving up to $N$ steps with each step size being $\delta\beta/2$. The latter choice is adopted in our simulations and decreases the Trotter error from $\mathcal{O}({\delta \beta}^2)$ to $\mathcal{O}({\delta \beta}^2/4)$.

\begin{figure}[ht]
    \centering
    \includegraphics[width = 0.75\columnwidth]{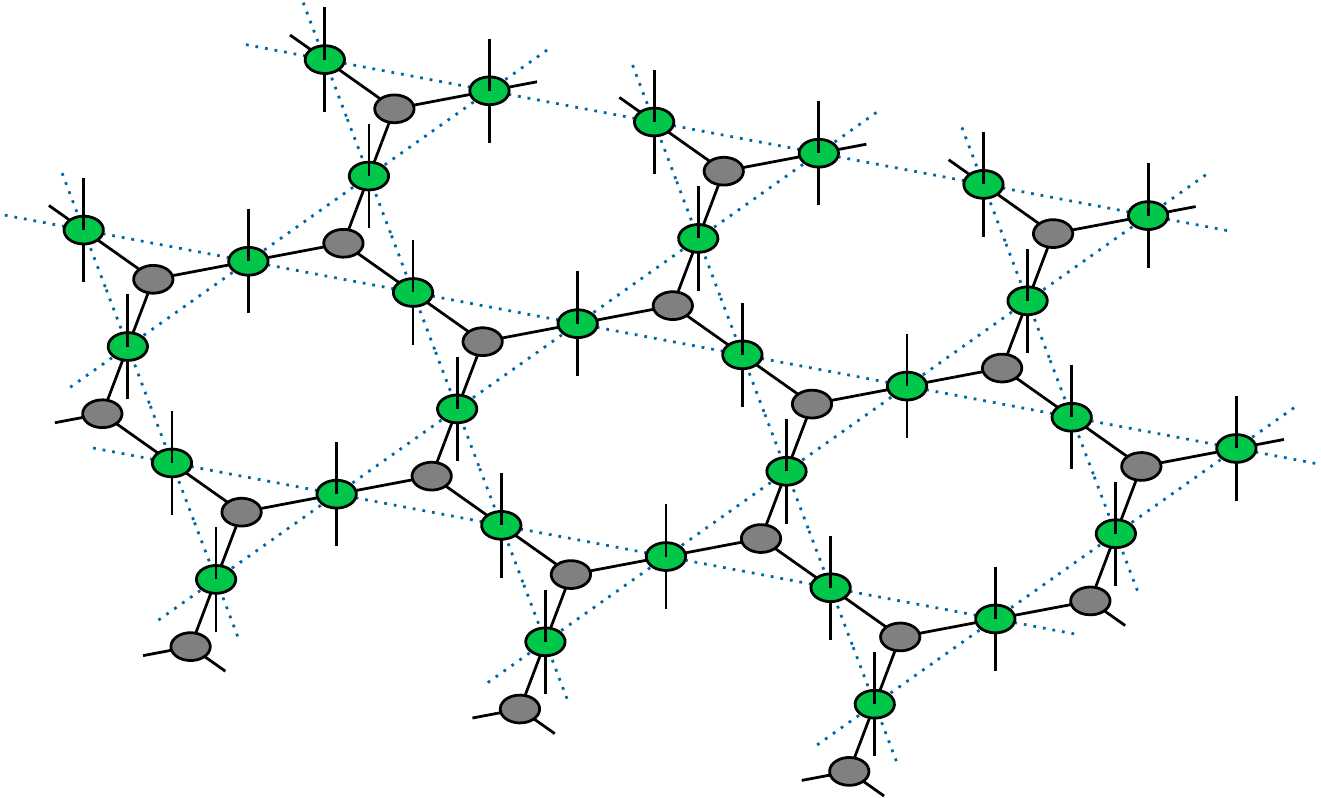}
    \caption{iPESO ansatz for the simulation of Gibbs states 
    on the Kagome lattice (shown in light blue).}
    \label{fig:iPESO_Ansatz_1}
\end{figure}

We will now introduce the tensor network representation of the Gibbs state: The \emph{infinite projected entangled simplex operator} (iPESO) shown in Fig.~\ref{fig:iPESO_Ansatz_1}. It is the operator version of the \emph{infinite projected entangled simplex state} (iPESS) proposed in Ref.~\cite{Xie2014}, applied to the simulation of thermal density matrices. In both tensor networks, the quantum correlations inherently present on the Kagome triangles are efficiently and accurately captured by exploiting the structure of its dual, the honeycomb lattice.
Green tensors represent the lattice sites of the Kagome lattice, with two physical indices for the density matrix (as opposed to a quantum state, for which tensors only have a single physical index). They are connected by purely virtual simplex tensors shown in grey. The accuracy with which the iPESO approximates the thermal density matrix is controlled by the bond dimension of the virtual bulk indices, denoted as $\chi_B$. It is important to note that $\chi_B$ needs to be chosen sufficiently large to prevent truncation effects in the ITDM. This leads to a minimal bond dimension of $p^2$, where $p$ is the dimension of the Hilbert space of the local physical degrees of freedom (a detailed explanation is given in the Appendix). For the final simulations of the targeted real material we choose the bond dimension such that the total truncation error is below $\sim 10^{-5}$, see Fig.~\ref{fig:truncationError_KagomeCompound_1}. Expectation values are then directly computed in the tensor network representation of the thermal state according to
\begin{align}
    \langle \hat O \rangle = { N_\rho}^{-1} 
    \text{Tr} \bigl\lbrack \rho(\beta/2)^\dagger \, \hat O \, \rho(\beta/2) \bigr\rbrack,
\end{align}
with a normalization factor $ N_\rho \coloneqq \textrm{Tr}\lbrack \rho(\beta/2)^\dagger \, \rho(\beta/2) \rbrack$. Expectation values can be computed by either using the simple update mean-field environment, or by a full \emph{corner transfer matrix renormalization group (CTMRG)} procedure~\cite{ctmrgNishino,ctmrgOrus2009,ctmrgOrus2012}, which captures quantum correlations more faithfully. For the latter, the environment bond dimension $\chi_E$ controls the approximations in the contraction of the infinite two-dimensional lattice. Details for both the simple update cooling and calculations of expectation values are presented in detail in the Appendix. The smallest possible unit cell of the iPESO consists of three lattice site tensors and two simplex tensors, as presented in Fig.~\ref{fig:iPESO_Ansatz_1}. Besides this structure, we also employ a nine-site unit cell in our numerical simulation. This is required to capture thermal states with larger structures that are not commensurate with three-site translational invariance.

\section{Models and results}
\label{modelresult}
\subsection{Kagome Heisenberg anti-ferromagnet}

\begin{figure}[ht]
    \centering
    \includegraphics[width = 1.0 \columnwidth]{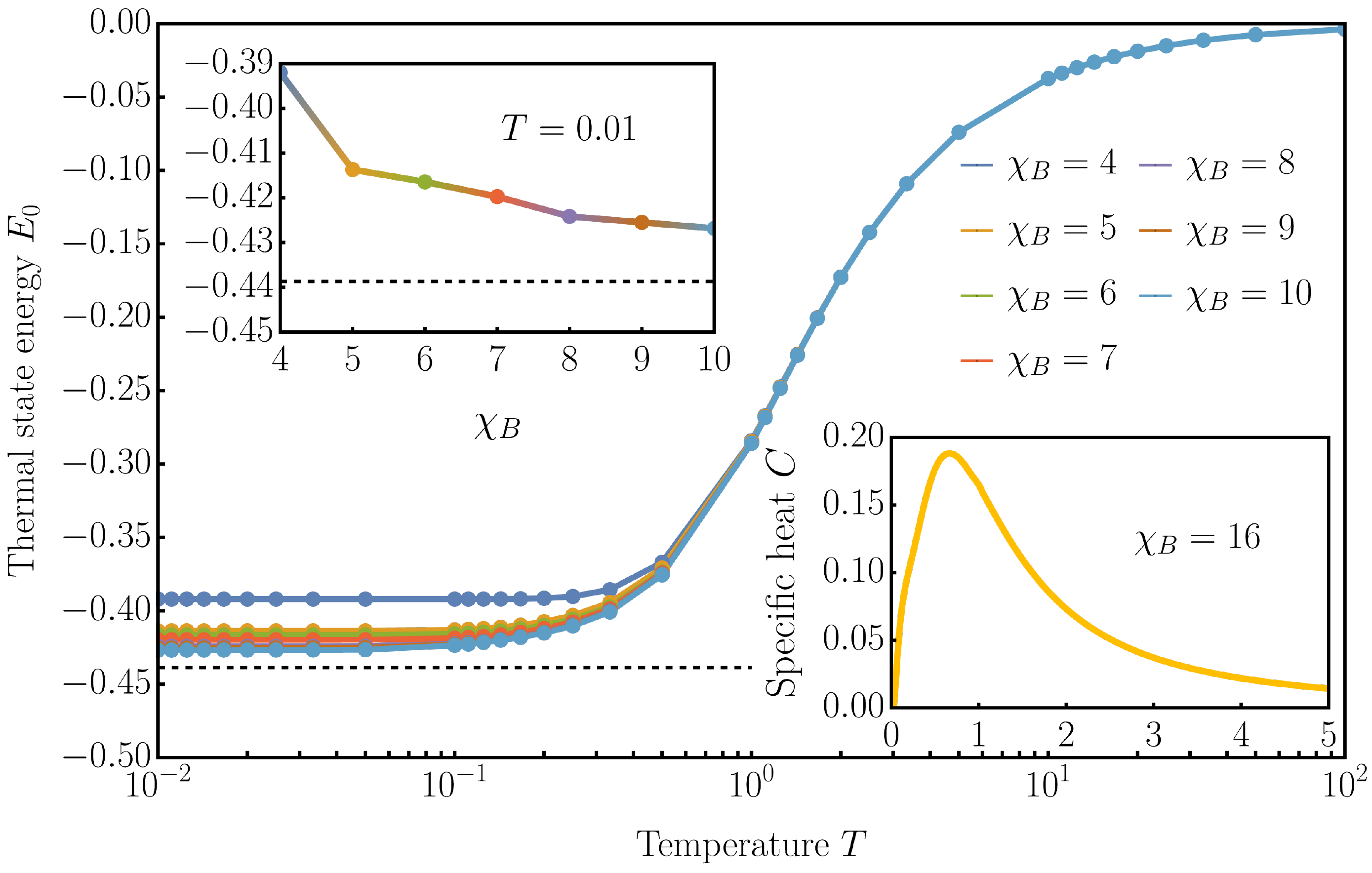}
    \caption{Thermal state energy for the spin-$1/2$ Heisenberg model on the Kagome lattice at various bulk bond dimensions $\chi_B$, using CTMRG environments. The dashed line corresponds to the $T = 0$ ground state energy of Ref.~\cite{Laeuchli2019}. (Top inset) Convergence of the thermal state energy with $\chi_B$ at $T = 0.01$. (Bottom inset) Specific heat $C$ at $\chi_B = 16$, using mean-field environments.}
    \label{fig:finalPlot_HeisenbergModel_1}
\end{figure}

The first application of the developed iPESO method is the finite temperature study of the frustrated \mbox{spin-$1/2$} Heisenberg anti-ferromagnet on the Kagome lattice, a paradigmatic model that has been a topic of intense study in the community~\cite{Sachdevkhafz2,Rankhafu1,dmrgkhafz2,Jiangkhafz2,Xiangkhaf,Gotzekhafz2,Iqbalkhafu1}. Its Hamiltonian is given by
\begin{align}
     H = J \sum_{\langle i,j \rangle} \vec S_i \cdot \vec S_j - h_z \sum_i S_i^z ,
    \label{eq:HeisenbergHamiltonian}
\end{align}
where $\vec S_i$ are spin-$1/2$ operators on site $i$ and $\langle i,j \rangle$ denotes nearest-neighbours in the underlying lattice, $h_z$ is a magnetic field applied along the $z$-axis. In the following, we employ the iPESO method to study the model at $J = 1.0$ over a large temperature range, choosing an infinitesimal temperature step \mbox{$\delta\beta = 10^{-3}$}. In the main panel of Fig.~\ref{fig:finalPlot_HeisenbergModel_1} we show the thermal state energy for a three-site iPESO ansatz at bulk bond dimensions up to $\chi_B = 10$. These results are computed using a CTMRG procedure with individual environment bond dimensions $\chi_E$ such that expectation values are well converged.
The energy of the thermal state approaches the ground state energy at $T = 0$ for low temperatures, as shown in the top inset. This, along with the vanishing magnetization when approaching the ground state (not shown here), indicates that the annealing procedure does not get stuck in local minima and flows towards the correct ground state as would have been obtained using ground state optimization. While the thermal state energies in Fig.~\ref{fig:finalPlot_HeisenbergModel_1} have been computed with CTMRG environments, we note that the accuracy is not affected while using the mean-field environment of the simple update. Therefore, we compute the \emph{heat capacity} $C \coloneqq \partial U / \partial T$ at a higher bulk bond dimension $\chi_B = 16$, using these environments. The result is shown in the bottom inset and matches previous finite temperature studies of the model~\cite{Chen2018_2}. In the analysis of the model and for the targeted real material we therefore restrict to mean-field environments, since CTMRG calculations are limited to inexpressively small environment bond dimensions in those cases.

\begin{figure}[ht]
    \centering
    \includegraphics[width = \columnwidth]{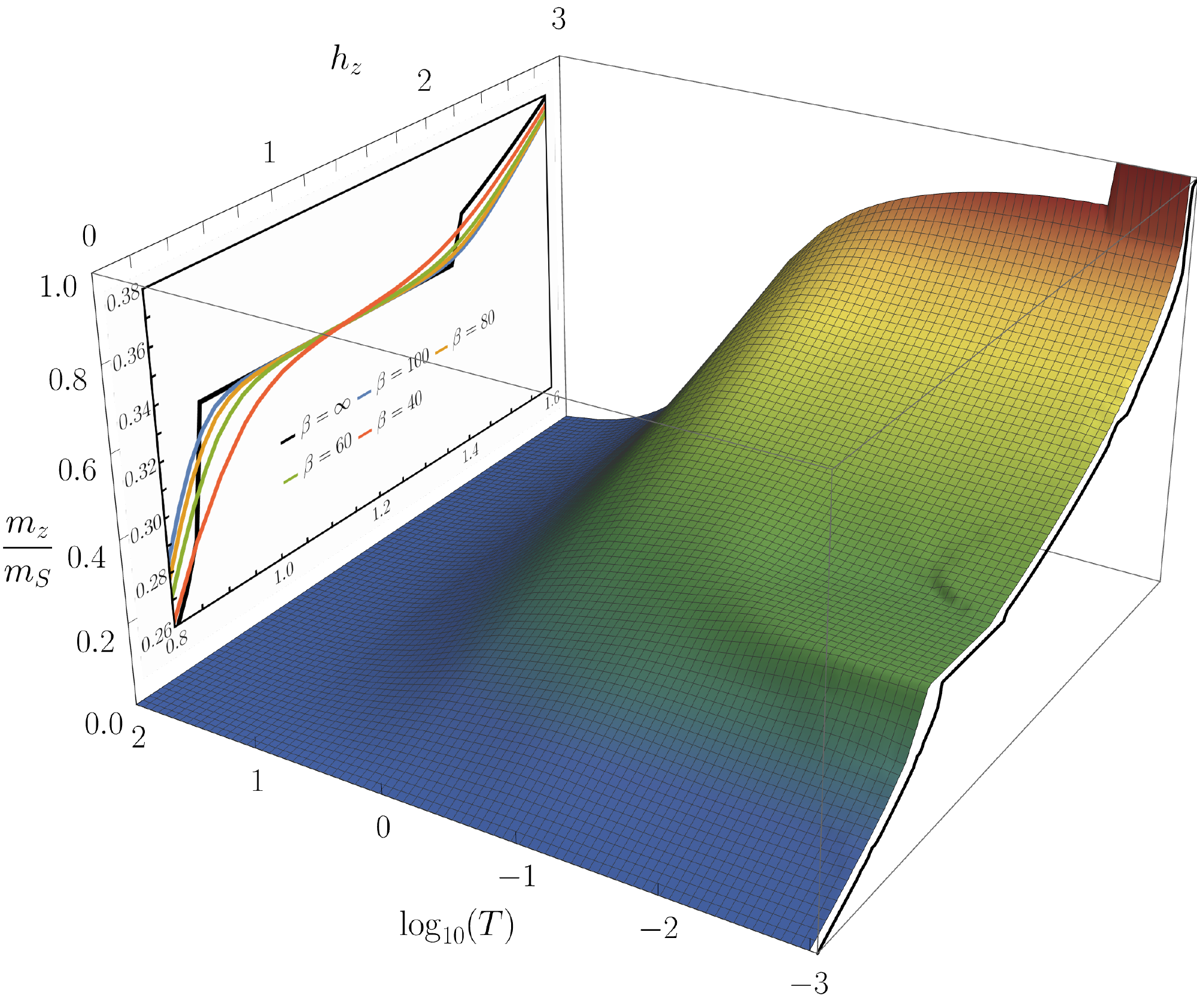}
    \caption{Magnetization of the Kagome Heisenberg anti-ferromagnet for various temperatures $T$ and magnetic fields $h_z$. Additionally, we show the $T = 0$ magnetization curve obtained via iPESS simulations in black. (Inset) Melting of the $m_z/m_S = 1/3$ plateau.}
    \label{fig:magneticPlateaus_Heisenberg}
\end{figure}

We further use our method to study the effect of temperature on the magnetization behaviour of the Heisenberg model in Eq.~\eqref{eq:HeisenbergHamiltonian}. It is known that four different magnetization plateaus at values $m_z/m_S = \lbrack 1/9, 1/3, 5/9, 7/9\rbrack$ of the saturation magnetization $m_S = 1/2$ appear at $T = 0$ upon tuning the magnetic field $h_z$~\cite{PicotspinS,KshetrimayumkagoXXZ}. 
We focus our study on the most prominent $m_z/m_S = 1/3$ magnetic plateau and classically simulate the Kagome Heisenberg model in a field for various temperatures in the range of $T \, \in \, \lbrack 0.001, 100 \rbrack$ using a nine-site iPESO at bond dimension $\chi_B = 12$. The results in Fig.~\ref{fig:magneticPlateaus_Heisenberg} are complemented with regular iPESS simulations at $T = 0$ and $\chi_B = 12$, where the magnetization plateaus appear most prominently.
Of all the $T=0$ magnetization plateaus, we find that only the $1/3$ plateau survives at finite temperature up to $T \sim 2\cdot 10^{-2}$, where $T$ is in units of the interaction strength $J$. Above this temperature, the plateau starts melting and disappears. A cross section of Fig.~\ref{fig:magneticPlateaus_Heisenberg} along constant temperature slices of $1/T = [\infty ,100, 80, 60, 40]$ reveals the melting of the plateau, shown as the inset on the face of the cube. We observe that the melting is stronger at the low-field end of the plateau, which is in agreement with a recent exact diagonalization study of the melting of the magnetization plateaus~\cite{Schlueter2022}. Our results serve as an important guide to the experimental study of the magnetization process of closely related real materials such as Herbertsmithite ZnCu$_3$(OH)$_6$Cl$_2$ and its relatives.

\subsection{Real material Ca$_{10}$Cr$_7$O$_{28}$}

\begin{figure}[ht]
    \centering
    \includegraphics[width = 0.65\columnwidth]{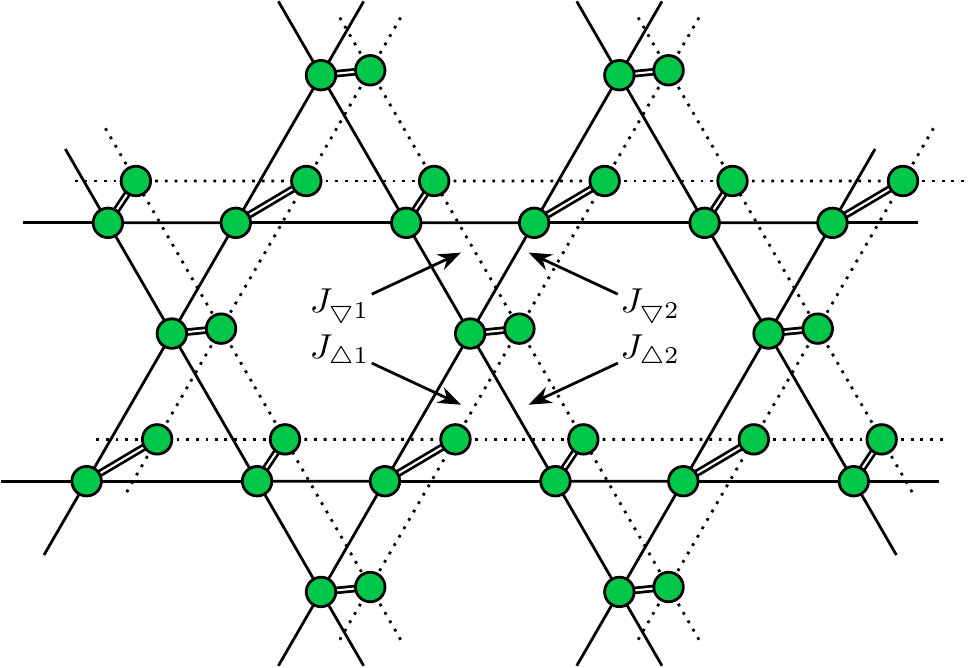}
    \caption{Breathing bilayer Kagome lattice with different couplings for the compound material Ca$_{10}$Cr$_7$O$_{28}$. The double bonds denote inter-layer coupling with coupling strength $J_\text{inter}$.}
    \label{fig:doubleLayerKagomeLattice_1}
\end{figure}

The material Ca$_{10}$Cr$_7$O$_{28}$ has a breathing bilayer Kagome structure, with alternating ferro- and anti-ferromagnetic Heisenberg interactions on neighbouring triangles, defined by
\begin{align}
    \begin{split}
         H &= \sum_{k = 1}^{2} \left\lbrack J_{\bigtriangledown k} \sum_{\langle i,j \rangle} \vec S_i \cdot \vec S_j + J_{\bigtriangleup k} \sum_{\langle i,j \rangle} \vec S_i \cdot \vec S_j \right\rbrack \\
        &+ J_\text{inter} \sum_{\langle i,j \rangle} \vec S_i \cdot \vec S_j - h_z \sum_i S_i^z
    \end{split},
    \label{eq:CaCrOHamiltonian}
\end{align}
where $J_{\bigtriangledown k}$ and $J_{\bigtriangleup k}$ are the intra-Kagome couplings in the two layers $k=1$ and $k=2$ while $J_\text{inter}$ denotes the coupling between the two layers. The lattice structure is shown in Fig.~\ref{fig:doubleLayerKagomeLattice_1}.
The different coupling parameters have been determined from neutron scattering experiments of the real material in Ref.~\cite{Balz2017}. These values (in meV) are $J_{\bigtriangledown 1} = +0.09(2)$, $J_{\bigtriangleup 1}=-0.27(3)$, $J_{\bigtriangledown 2} = -0.76(5)$, $J_{\bigtriangleup 2} = +0.11(3)$ and $J_\text{inter} = -0.08(4)$, {where the numbers in round brackets indicate the uncertainties}.

The bilayer Kagome structure can be mapped to a single layer by combining the two spins in the different layers to a single physical site, so that a regular iPESO ansatz with an enlarged local physical dimension of $d=4$ can be used. The large physical dimension of the system also means that the minimum bond dimension required for accurate simulations of the material needs to be sufficiently high ($\chi_B \geq 16$). This becomes a bottleneck in computing expectation values using CTMRG routines. For this reason, we have adopted the mean-field environment calculation which takes into account the quantum correlations within a cluster. We find that even with such approximations in computing the magnetization and heat capacity, our results are compatible with the experimental data and provide novel insights into the earlier discrepancy between theory and experiment data as we show below.

\subsubsection{Magnetization behaviour}

The magnetic properties of this real material have been investigated previously at zero temperature using tensor networks and compared to experimental measurements by some of the current authors~\cite{Kshetrimayummaterial}. While results at small magnetic fields were in good agreement, the magnetization curve has shown a significant discrepancy between simulation and experiment at large values of the field. Such a discrepancy has also been observed when comparing the experimental data to theoretical mean-field calculations~\cite{Balz2017}. Experimental data show that the magnetization of this material increases rapidly for small external magnetic fields up to $1$ or $\SI{2}{\tesla}$ above which the slope flattens and saturation is achieved for a field value of $\approx \SI{12}{\tesla}$. In contrast, our previous tensor network simulation predicted saturation at a much smaller value of the external magnetic field of $\approx \SI{1}{\tesla}$. This theoretical investigation, while quantum, has ignored thermal fluctuations. For comparison, molecular dynamics simulation which take temperature effects into account result in a better agreement~\cite{MD}, however the quantum fluctuations are ignored by this method. Our present technique encompasses both the quantum properties of the material as well as the effect of finite temperature.

\begin{figure}[ht]
    \centering
    \includegraphics[width = 1.0\columnwidth]{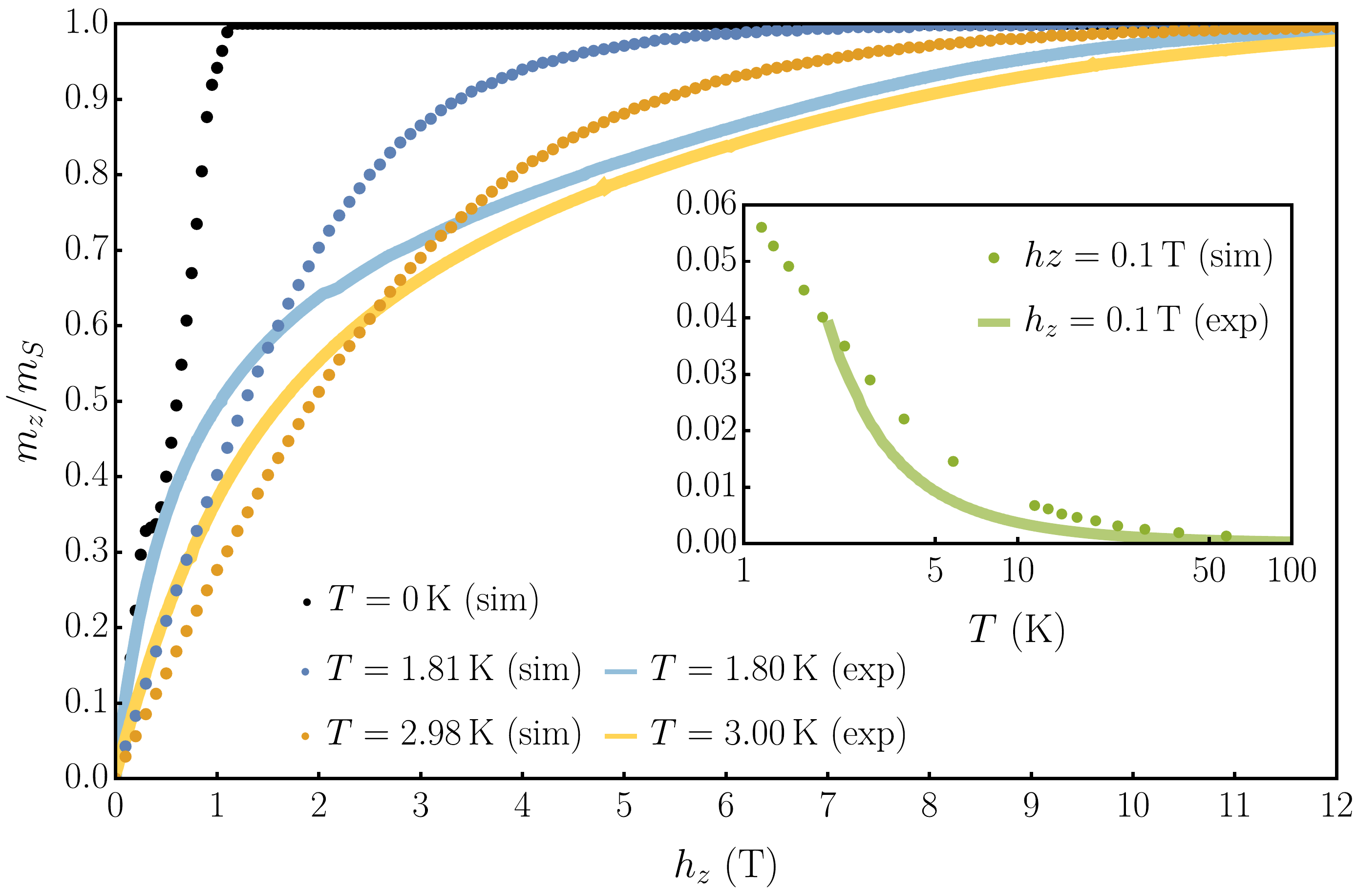}
    \caption{Magnetization curve of Ca$_{10}$Cr$_7$O$_{28}$ for the Hamiltonian given in Eq.~\eqref{eq:CaCrOHamiltonian} and magnetic fields between $\SI{0}{\tesla}$ and $\SI{12}{\tesla}$ at different temperatures. With decreasing temperature, the slope increases and the magnetization saturates earlier. The inset shows the magnetization at fixed field of $h_z = \SI{0.1}{\tesla}$ over temperature.}
    \label{fig:magetizationZ_KagomeComp_Lx_1_Ly_3}
\end{figure}

We now investigate the quantum material Ca$_{10}$Cr$_7$O$_{28}$ in the presence of magnetic fields between $\SI{0}{\tesla}$ and $\SI{12}{\tesla}$ at various temperatures. The TN simulations are done with a nine-site iPESO at bond dimension $\chi_B = 30$ and with $\delta\beta = 10^{-2}$, results are shown in Fig.~\ref{fig:magetizationZ_KagomeComp_Lx_1_Ly_3}.
Our theoretical results are plotted for temperatures \mbox{$T = \SI{1.81}{\kelvin}$} and \mbox{$T = \SI{2.98}{\kelvin}$}. This is then compared against the experimental data measured at \mbox{$T = \SI{1.8}{\kelvin}$} and \mbox{$T = \SI{3.0}{\kelvin}$}. The conversion factors between the theoretical calculations and experiment are shown in the Appendix. From the plots in Fig.~\ref{fig:magetizationZ_KagomeComp_Lx_1_Ly_3}, we see that as we increase the temperature, the field value at which the magnetization saturates becomes larger and approaches the experimental findings. Overall, we find that the magnetization curve shows a strong dependence on the temperature and the saturation sets in quicker for low $T$. This seems to indicate that the earlier discrepancy between theory and experimental data has been largely due to neglecting finite temperature effects in the theory simulations. For comparison, we show the $T = \SI{0}{\kelvin}$ magnetization curve obtained with a nine-site iPESS at $\chi_B = 24$ with mean-field environments.

Inelastic neutron scattering has previously revealed that the spin liquid ground state of this material is destroyed by a magnetic field of $\SI{1}{\tesla}$~\cite{Balz2017}. However, heat capacity measurements could show that magnetic fields of up to $\SI{0.5}{\tesla}$ leave the spin liquid ground state intact as indicated by featureless $C/T$ curves~\cite{Balz2017}. To check that the effect of small fields is correctly captured by our model, we have computed the magnetization as a function of temperature at fixed field strength $h_z = \SI{0.1}{\tesla}$ and contrasted it with measured experimental data. The comparison is shown in the inset of Fig.~\ref{fig:magetizationZ_KagomeComp_Lx_1_Ly_3}, and the model indeed reproduces the experimental curve without any anomalies that would indicate a phase transition into a magnetically ordered ground state.

\subsubsection{Heat capacity and entropy}

\begin{figure}[ht]
    \centering
    \includegraphics[width = 1.0\columnwidth]{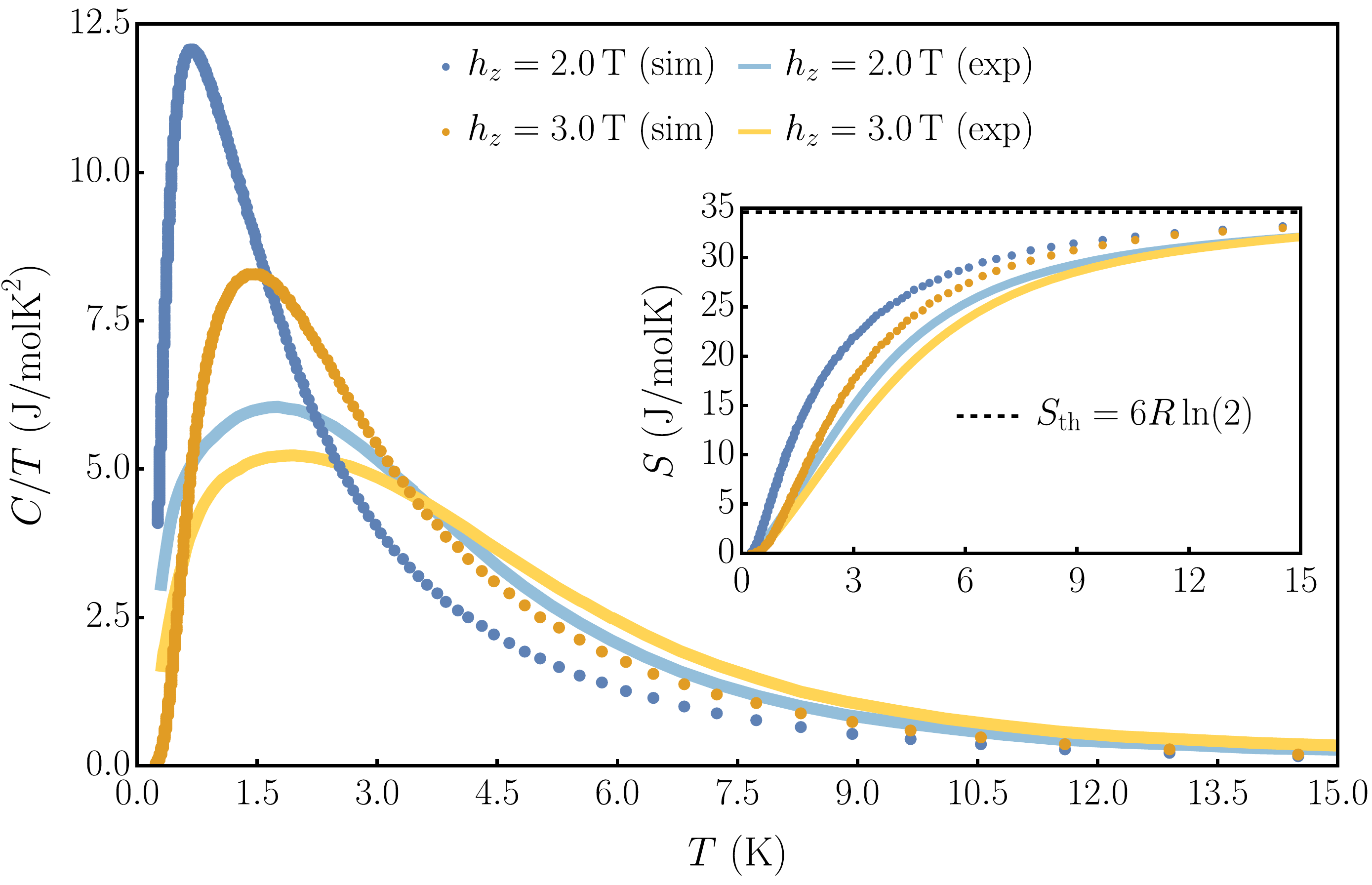}
    \caption{Magnetic heat capacity $C/T$ as a function of temperature for different strengths of the magnetic fields $h_z$. Integrating this quantity over $T$ yields the entropy, which accurately approaches the theoretical value of $S_\textrm{th} = 6R\ln(2)$, with $R$ the ideal gas constant.}
    \label{fig:heatCapacity_Lx_1_Ly_3_dBeta_1.00e-02}
\end{figure}

Finally, we compute the magnetic heat capacity from the thermal state energy $U$ according to $C \coloneqq \partial U /\partial T$ for two different values of the magnetic field $h_z = \SI{2.0}{\tesla}$ and $h_z = \SI{3.0}{\tesla}$ and compare it with the experimental data. Results are shown in Fig.~\ref{fig:heatCapacity_Lx_1_Ly_3_dBeta_1.00e-02}. The conversion factors between the theoretical calculations and experiment are again shown in the Appendix.
As well as the heat capacity data, we have also computed the thermodynamic entropy by integrating the heat capacity. This is shown in the inset of Fig.~\ref{fig:heatCapacity_Lx_1_Ly_3_dBeta_1.00e-02}. We find good agreement between our theoretical predictions and experimental data. 

The heat capacity of Ca$_{10}$Cr$_7$O$_{28}$ at intermediate fields is characterized by a broad and smooth peak of a Schottky anomaly due to the excitations that become gapped by the magnetic field. The position of this peak shifts to higher temperature with increasing field both in the model and the experimental data. This is in qualitative agreement with an increasing gap due to the Zeeman term in the Hamiltonian, see Eq.~\ref{eq:CaCrOHamiltonian}. Integrating $C/T$ to obtain the magnetic entropy shows that the model does well in capturing the total possible entropy for spin-1/2 over the temperature range up to $\SI{15}{\kelvin}$.

\section{Conclusions and outlook}

In this work, we have presented a two-dimensional tensor network algorithm for studying finite temperature properties for the highly challenging realm of frustrated systems and two-dimensional quantum materials. We achieve this by introducing the infinite projected entangled simplex operator
algorithm. Our algorithm explicitly preserves the positive semi-definiteness of the Gibbs state represented by the iPESO. We use our technique to benchmark against finite temperature properties of the well-known, paradigmatic model of the spin-1/2 Kagome Heisenberg anti-ferromagnet and obtain very competitive state-of-the art results for the thermal state energy and heat capacity. We also study the melting of the magnetization plateaus of this model in the presence of external magnetic field at finite temperature. By focusing on the most prominent $1/3$ plateau, we find that it starts melting and disappears at temperature $T \sim 2\cdot 10^{-2}$. Moreover, the plateau starts melting from the lower end of the field, an observation that was also made recently in an independent exact diagonalization study~\cite{Schlueter2022}.

Finally, we have investigated the finite temperature properties of the quantum material Ca$_{10}$Cr$_7$O$_{28}$ using our tensor network technique. This is particularly important due to a recent discrepancy in the magnetization process predicted by theoretical simulations compared to experimental findings. As a first theoretical study of this real material that includes both quantum correlations and finite temperature effects, we find a strong temperature dependence of the magnetization curve of this material in the presence of an external magnetic field. We find that on systematically increasing the temperature, our theoretical simulations approach the experimental data which was collected at finite temperature. We provide a direct comparison of the theoretical magnetization data with the experimental data at $T = \SI{1.8}{\kelvin}$ and $T = \SI{3.0}{\kelvin}$ and find them to be in surprisingly good but not quite perfect agreement. We also computed the magnetic heat capacity ($C/T$) as a function of temperature at different field strengths $h_z = \SI{2.0}{\tesla}$ and $h_z = \SI{3.0}{\tesla}$ as well as the entropy $S$. For all these quantities, we find good agreement with the experimental data. 

One can argue that the agreement is striking, given that the Hamiltonian in Eq.~(\ref{eq:CaCrOHamiltonian}) has only been recovered by neutron scattering techniques to finite precision considering five Heisenberg interactions, while Dzyaloshinskii-Moriya interactions have been excluded. Furthermore, there are truncation errors in the classical simulation. One can argue that the present analysis allows to cross-benchmark quantum experiments with classical simulations. The findings can also be seen as an invitation, however, to use high-precision tools of \emph{Hamiltonian learning} to better identify the actual underlying microscopic Hamiltonian, given data from Gibbs states~\cite{GibbsLearning,GibbsLearning2}, possibly even based on tensor networks akin the approach taken in Ref.~\cite{HamLearning}. These steps would further contribute to an engineering perspective of studying realistic strongly correlated quantum materials with tensor networks.

We believe our work to be an important step towards bridging the gap between theoretical simulations and experimental studies of quantum materials. It would be straightforward to extend our algorithm to other lattices and geometries that may suit other quantum materials. By incorporating both quantum correlations and finite temperature effects, we have now made direct comparison between experimental and theoretical data possible.

\section{Acknowledgements}

The authors are thankful for discussions with Ji-Yao Chen, Dante Kennes, Corinna Kollath, David Luitz, Jan Naumann, Román Orús, Matteo Rizzi and Anne-Maria Visuri. The authors would like to thank the HPC Service of ZEDAT, Freie Universität Berlin, for computing time~\cite{Bennett2020}. The FUB team acknowledges funding by the Deutsche Forschungsgemeinschaft (CRC 183 on `Entangled states of matter' and FOR 2724 on `Thermal machines in the quantum world'), the Helmholtz Association, and the BMBF (MUNIQC-ATOMS), for which this work constitutes method development. B.~L.~acknowledges the support of Deutsche Forschungsgemeinschaft through project B06 of SFB 1143 on `Correlated magnetism: From frustration to topology' (ID 247310070). 

\bigskip

%

\begin{appendix}

\newpage
\section{Details of the tensor network algorithm}
\label{app:TensorNetworkDetails}

\subsection{Simple update}
\label{app:SimpleUpdate}

\begin{figure}[ht]
    \centering
    \includegraphics[width = 0.9\columnwidth]{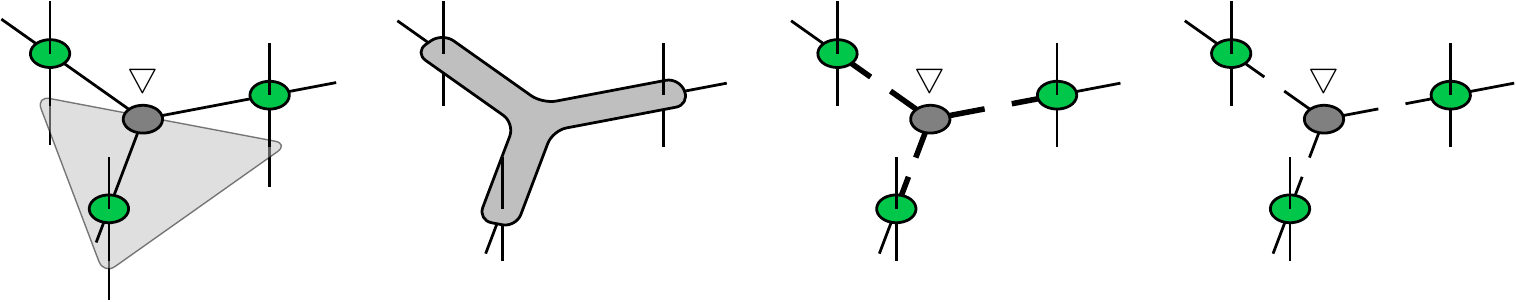}
    \caption{Simple update step for the (imaginary) time evolution of the iPESO ansatz. 
    After applying the three-body gate to the physical indices, 
    a truncated higher-order SVD is used to decompose the nine-index tensor and restore the lattice and simplex tensors.}
    \label{fig:iPESO_simpleUpdate_thermalDensityMatrix_1}
\end{figure}

The simple update describes an efficient, yet approximate scheme to do the annealing respectively the imaginary time evolution of the initial density matrix. It essentially implements the evolution in Eq.~\eqref{eq:STFirstOrder} together with a local truncation to keep the bulk bond dimension fixed. Without the loss of generality, we consider a Hamiltonian with local interactions in the form of
\begin{align}
     H = H_{\bigtriangledown} + H_{\bigtriangleup},
\end{align}
where $H_{\bigtriangledown}$ and $H_{\bigtriangleup}$ are three-spin interactions on the two types of triangles of the Kagome lattice, respectively. Making use of a first-order 
\emph{Suzuki-Trotter decomposition}, the imaginary time evolution to evolve the thermal density matrix $\rho(\beta) \rightarrow \rho(\beta + \delta\beta)$ can be approximated by applying the operator
\begin{align}
    \begin{split}
        U(\delta\beta) &= \mathrm e^{- \delta\beta   H_{\bigtriangledown}} \mathrm e^{- \delta\beta H_{\bigtriangleup}} + O(\delta\beta^2) \\
    & \approx U_{\bigtriangledown}(\delta\beta) U_{\bigtriangleup}(\delta\beta)
    \end{split}
    \label{eq:STFirstOrder}
\end{align}
to both three-site configurations in the tensor network. In Fig.~\ref{fig:iPESO_simpleUpdate_thermalDensityMatrix_1} we illustrate the evolution of the iPESO with the three-body gate $U_{\bigtriangledown}(\delta\beta)$. This step involves three lattice tensors, as well as the simplex tensor $\bigtriangledown$.
After the gate has been contracted with the tensors, a \emph{higher-order singular value decomposition} (HOSVD) with subsequent truncation is used to separate the network back into simplex and lattice tensors. Since the truncation is based only on the singular values for the three indices, it is purely local. In a similar fashion, the simplex $\bigtriangleup$ is updated alongside the three lattice site tensors by applying the three-body gate $U_{\bigtriangleup}(\delta\beta)$.
After both steps have been performed, we obtain the thermal density matrix $\rho(\beta + \delta\beta)$, represented by a three-site iPESO. This process is repeated for a fixed number of steps, such that the final thermal density matrix represents the quantum system at the desired (inverse) temperature. Naturally, this can be extended to Hamiltonians with less or more than three-site interactions. Additionally, physical symmetries of the Hamiltonian (like $U(1)$ or $SU(2)$) can be readily directly incorporated, exploiting symmetry-preserving tensors~\cite{Singh2010, Schmoll2020}.

\subsection{Bond dimension considerations}
\label{app:BondDimensionConsiderations}

\begin{figure}[ht]
    \centering
    \includegraphics[width = 0.9\columnwidth]{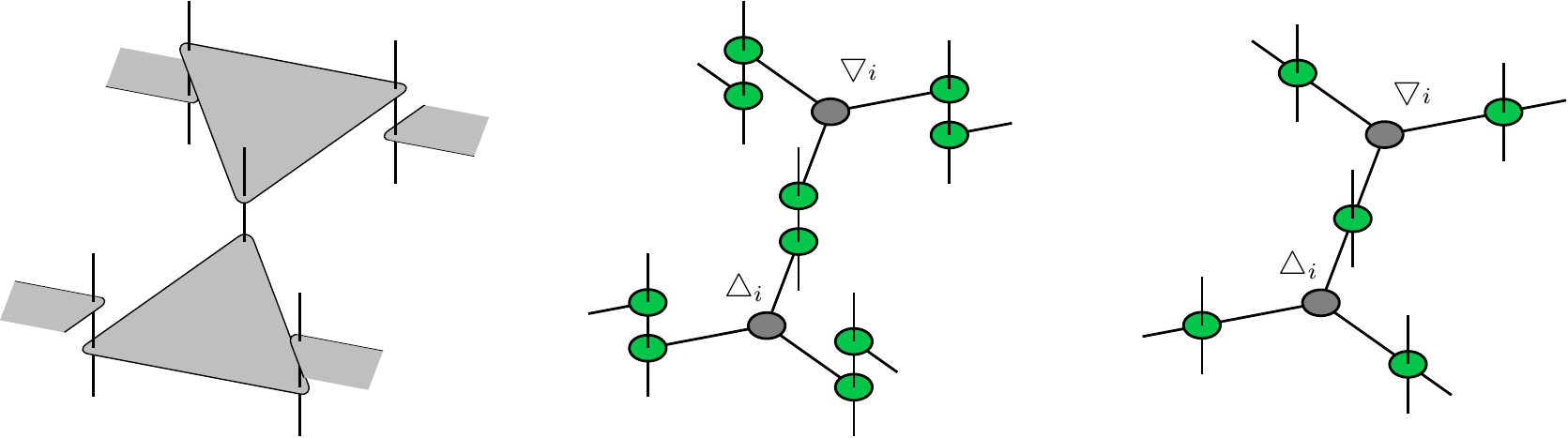}
    \caption{Construction of the infinitesimal thermal density operator $\rho(\delta\beta)$ by a decomposition of the Trotter gates (see Eq.~\eqref{eq:STFirstOrder}). An exact representation (apart from the inevitable Trotter error) can only be achieved without truncating the virtual bulk bond dimension, which is therefore at least $p^2$.}
    \label{fig:iPESO_Multiplication_2}
\end{figure}

As it is common in tensor network applications, the bond dimension controls the precision of the simulations. Here, we aim at presenting a discussion of the minimal bond dimensions required in order to obtain meaningful results. The infinitesimal thermal density matrix
\begin{align}
    \rho(\delta\beta) = \prod_{\langle i,j,k \rangle} \mathrm{e}^{-\delta\beta H_{i,j,k}} + \mathcal O(\delta\beta^2),
    \label{eq:infintesimalDensityMatrix}
\end{align}
here to first order in the Suzuki-Trotter decomposition, can be constructed by applying the Trotterized Hamiltonian gates $\exp(-\delta\beta H_{i,j,k})$ onto the infinite temperature Gibbs state $\rho(\beta = 0)$, as shown in Fig.~\ref{fig:iPESO_Multiplication_2} (the infinite temperature state is simply a tensor product of identity matrices).
An accurate representation (within the inevitable Trotter error) of this state is only possible if the resulting iPESO tensors are not truncated. Since the infinite temperature state has a bond dimension of unity, the infinitesimal thermal density matrix necessarily has bond dimension $p^2$, where $p$ is the physical dimension of the system. Naturally, cooling the state down to lower temperatures can only produce meaningful results, if the bond dimension is larger than the minimally necessary one.

\subsection{Effective environments and expectation values}
\label{app:EffectiveEnvironments}

\begin{figure}[ht]
    \centering
    \includegraphics[scale = 0.7]{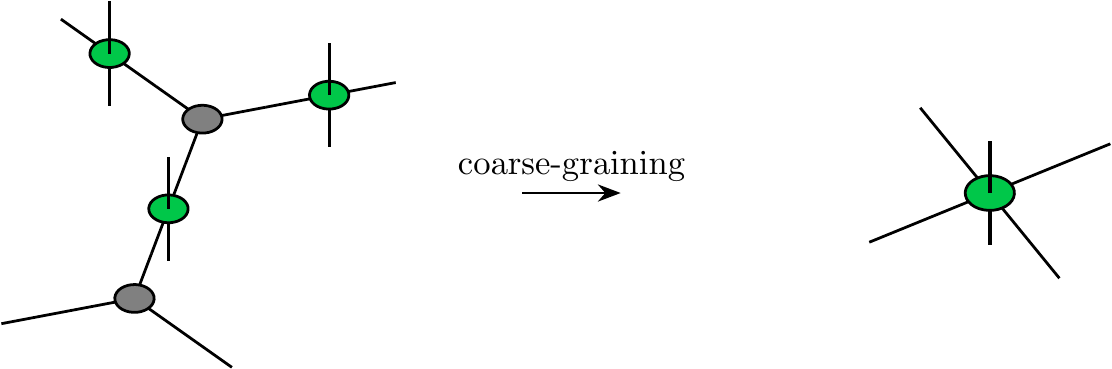}
    \caption{Coarse-graining of a three-site iPESO into a single-site iPEPO tensor. For larger iPESO unit cells the resulting iPEPO network will have a larger unit cell, too.}
    \label{fig:iPESO_CoarseGraining_1}
\end{figure}

\begin{figure}[ht]
    \centering
    \includegraphics[width = 0.95\columnwidth]{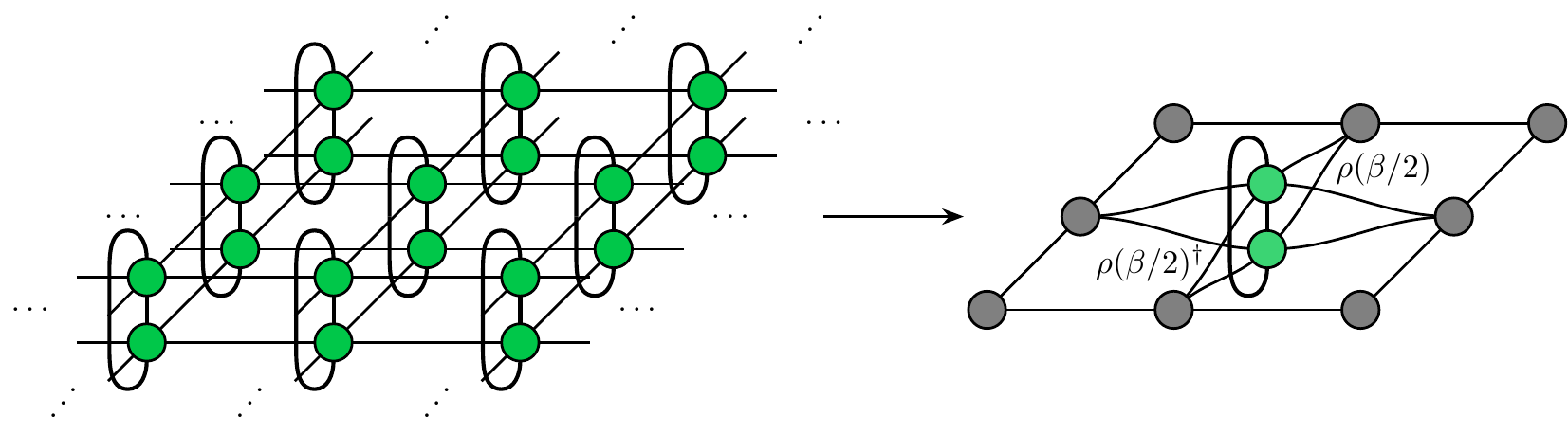}
    \caption{A directional CTMRG routine is used to approximate the contraction of the infinite square lattice by a set of fixed-point environment tensors, denoted in grey.}
    \label{fig:CTMRG_Method_1}
\end{figure}

In order to evaluate physical observables and compute expectation values accurately, the infinite two-dimensional iPESO tensor network needs to be contracted. It is known that this task cannot be performed exactly classically 
efficiently both in worst case and average case complexity~\cite{PhysRevLett.98.140506,PhysRevResearch.2.013010}, without an exponential increase in computation time, so that approximate methods must be employed. Here we utilize the so-called \emph{corner transfer matrix renormalization group} (CTMRG) to compute the effective environment tensors for every lattice site. To this end, we coarse-grain the iPESO network to an iPEPO network, the operator form of the famous \emph{infinite projected entangled pair state} (iPEPS), as visualized in Fig.~\ref{fig:iPESO_CoarseGraining_1}.
After coarse-graining, the environment surrounding each local thermal density matrix can be conveniently computed using a standard CTMRG procedure. To this end, the contraction of the infinite square lattice is approximated by a set of fix-point environment tensors, as shown in Fig.~\ref{fig:CTMRG_Method_1}.
This enables both accurate calculations of expectation values and would be essential in devising a sophisticated update procedure that includes all quantum correlations in the system \textemdash\ the so-called \emph{full update}~\cite{Phien2015, Czarnik2019}. Since the two physical indices are traced over, the procedure is a straightforward extension of a regular CTMRG routine for a two-dimensional iPEPS wave function. In order to ensure that the thermal density matrix is reflected by a positive semi-definite operator, a double-layer approach is taken in contrast to the original proposal in Ref.~\cite{Kshetrimayum2019}.

\subsection{Truncation effects in the simple update}
\label{app:TruncationEffects}

\begin{figure}[ht]
    \centering
    \includegraphics[width = 1.0\columnwidth]{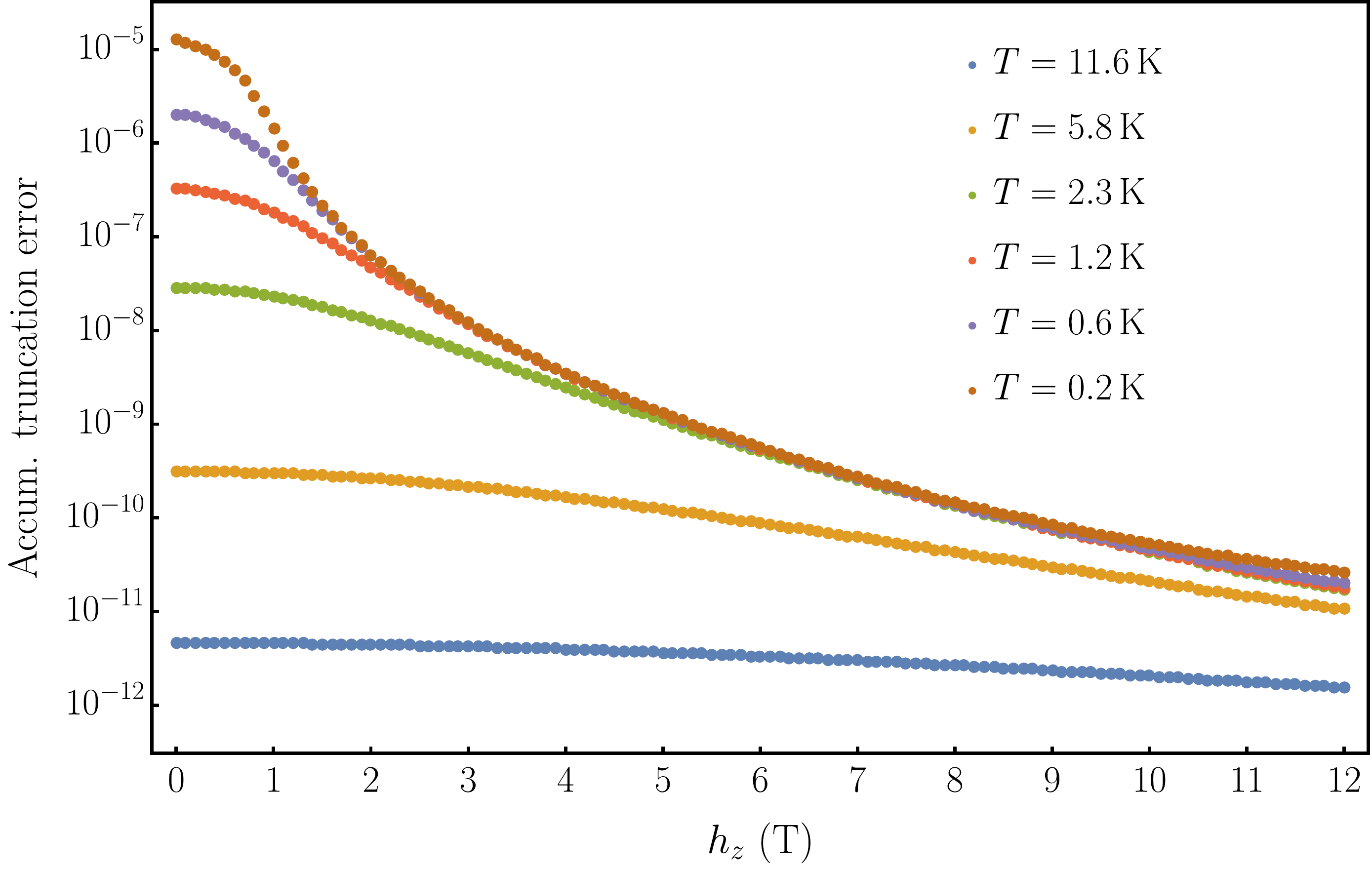}
    \caption{Accumulated truncation error in the simple update annealing for the bilayer Kagome compound Ca$_{10}$Cr$_7$O$_{28}$ at \mbox{$\chi_B = 30$} and for $\delta\beta = 10^{-2}$.}
    \label{fig:truncationError_KagomeCompound_1}
\end{figure}

The annealing scheme adopted in this study is based on the simple update, which requires truncations in order to keep the bulk bond dimension $\chi_B$ constant. Moreover, the choice of the infinitesimal cooling step $\delta\beta$ controls the unavoidable error in the Trotterization, and the number of annealing steps which include a truncation. For the iPESO simulations of Ca$_{10}$Cr$_7$O$_{28}$ we choose a step size of $\delta\beta = 10^{-2}$, which leads to the accumulated truncation errors shown in Fig.~\ref{fig:truncationError_KagomeCompound_1}, for several temperatures.

The accumulated truncation error is given by the sum of the discarded weights of all simple update annealing steps. The discarded weight is the sum of the discarded squared singular values in the \emph{singular value decomposition} (SVD)~\cite{Schollwoeck2011}. Naturally, the truncation error decreases with increasing magnetic field, since the thermal states become closer to a reduced density matrix of a product state. Moreover, it increases with decreasing temperature, because more cooling steps are necessary to reach lower temperatures. In general, the large bond dimension of $\chi_B = 30$ keeps the accumulated truncation error low enough for our results to be meaningful down to the lowest temperatures we consider.

\section{Conversion between experiment and simulation}
\label{app:ConversionExpSim}

In order to compare the simulated tensor network data with measured experimental data, we need to apply the correct conversion factors. Since the coupling constants in the Hamiltonian are given in units of $\si{\milli\eV}$ and we set $k_B = 1$, the temperature $T_\text{sim}$ is in $\si{\milli\eV}$, too. The proper conversion to $\si{\kelvin}$ is, 
therefore, given by
\begin{align}
    \frac{T_\text{exp}}{T_\text{sim}} = \frac{\SI{1}{\milli\eV}}{k_B} = \frac{\SI{1.602e-22}{\joule}}{k_B} \approx \SI{11.6}{\kelvin}.
\end{align}
Furthermore, we need to convert the heat capacity $C$ between simulated and measured data. The tensor network data is given per spin in units of $\si{\milli\eV\per\kelvin}$. In order to convert it, a factor of
\begin{align}
    \frac{C_\text{exp}}{C_\text{sim}} = 6 \cdot \SI{1.602e-22}{\joule} \cdot N_A \approx \SI{578.8}{\joule\per\mol}
\end{align}
with $N_A = \SI{6.022e23}{\per\mol}$ the Avogadro constant, is required. The additional factor of six stems from the fact, that one formula unit of Ca$_{10}$Cr$_7$O$_{28}$ has six spin-$1/2$ chromium ions.

\end{appendix}

\end{document}